\documentstyle[12pt,aasms4,flushrt]{article}

\newcommand{\be}{\begin{equation}}
\newcommand{\ee}{\end{equation}}
\newcommand{\tbar}{ {\langle t \rangle}} 

\begin{document}

\title{CONSTRAINTS ON THE INTERGALACTIC TRANSPORT OF COSMIC RAYS}

\author{Fred C. Adams, Katherine Freese, Gregory Laughlin, 
Gregory Tarl{\'e}, and Nathan Schwadron} 

\affil{Physics Department, University of Michigan \\
	Ann Arbor, MI 48109-1120}

\date{April 1997} 

\begin{abstract} Motivated by recent experimental proposals to search
for extragalactic cosmic rays (including anti-matter from distant
galaxies), we study particle propagation through the intergalactic
medium (IGM). We first use estimates of the magnetic field strength
between galaxies to constrain the mean free path for diffusion of
particles through the IGM. We then develop a simple analytic model 
to describe the diffusion of cosmic rays.  Given the current age 
of galaxies, our results indicate that, in reasonable models, a 
completely negligible number of particles can enter our Galaxy from
distances greater than $\sim 100$ Mpc for relatively low energies 
($E$ $< 10^6$ GeV/n).  We also find that particle destruction in
galaxies along the diffusion path produces an exponential suppression
of the possible flux of extragalactic cosmic rays.  Finally, we use
gamma ray constraints to argue that the distance to any hypothetical
domains of anti-matter must be roughly comparable to the horizon
scale.  \end{abstract}

\keywords{ISM: Cosmic Rays -- Elementary Particles -- 
Cosmology: Theory -- Intergalactic Medium} 

\section{Introduction}		\label{sec:intro}

Cosmic rays of extragalactic origin can potentially provide an
important probe of our universe. However, the propagation of 
cosmic rays is highly constrained by magnetic fields, both in the 
interstellar medium and in the intergalactic medium (IGM).  In this
paper, we outline the basic issues involved in cosmic ray propagation
through the IGM and into the Galaxy.  In particular, we show that the
total distance traveled by cosmic rays during the age of the universe
is severely limited. These results have implications for recently
proposed experiments to detect anti-matter and ordinary cosmic rays
from external galaxies (see also Ormes et al. 1997). 

The framework for this paper can be summarized as follows.
[1] In the intergalactic medium, we assume that cosmic rays diffuse
with a characteristic mean free path.  By varying the mean free path,
one can investigate particle propagation in a large number of physical
scenarios.  [2] Particles have a limited accessibility to individual
galaxies.  This accessibility may depend upon galactic winds and/or
magnetic barriers.  [3] The propagation of cosmic rays depends on the
particle energies.  For low energy cosmic rays, $E \sim 1 - 10$ 
GeV/n, the particles are likely to follow magnetic field lines.  For
high energy particles, $E \gg 10^{18}$ eV/n, the particles no
longer follow field lines and random walk through space.  For
intermediate energies, the particle propagation problem is much harder
to describe.  In any case, the dependence on cosmic ray energy should
be kept in mind throughout this discussion.

We note that the global magnetic field structure of the universe
remains uncertain. The goal of the paper is to explore the optimal
case for transport of particles to our Galaxy from long distances.
Even with the most optimistic plausible choice of parameters, we find
that the particles cannot propagate farther than a few hundred Mpc
during the lifetime of the universe.  Our analysis using this
framework will be useful for evaluating the feasibility of various
detection strategies for future experiments.

This paper is organized as follows. We first estimate the magnetic
field strength and discuss the corresponding constraints on the mean
free path for cosmic rays (\S 2). In \S 3, we develop a simple
analytic model to describe the self-similar diffusion of cosmic ray
particles through the IGM; we use this model to estimate the
abundances of extragalactic cosmic rays and anti-matter. We also use
gamma ray constraints to estimate the distance to hypothetical domains
of anti-matter. We conclude in \S 4 with a summary and discussion of
our results.

\section{Magnetic Field Limits and Mean Free Paths} \label{sec:magnetic}

In this section, we discuss the field strength and the coherence
length for magnetic fields in the IGM.  Magnetic fields greatly
influence the motion of charged particles such as cosmic rays,
provided that the particle pressure is small compared to the magnetic
pressure (as we assume here).  We can consider two different regimes
of interest.  For cosmic rays of sufficiently low energy, the magnetic
gryo radius $r_B$ is small compared to the coherence length of the
magnetic field; in this limit, particles tend to follow the field
lines and the magnetic field geometry determines the paths taken by
the particles.  In the opposite limit of high energy cosmic rays, the
gyro radius is large compared to the coherence length; in this case,
the particles exhibit a random walk with a mean free path comparable
to the gyro radius, i.e., $\ell$ $\sim r_B$.  The magnetic field 
strength thus determines the energy boundary between low energy 
and high energy particles.  In addition, for high energy particles, 
the field strength determines, in part, the mean free path. 

We thus need an estimate of the magnetic field strength in the IGM.
We first consider the simplest theoretical considerations.  Given that
the galactic field strength is $B_{\rm gal} \approx 1-3 \mu$G (Heiles
1976), we can estimate the magnetic field strength $B_{IGM}$ between
galaxies in the following two ways:

[1] We consider the flux freezing approximation, in which the galaxy 
formed from a much larger region with a magnetic flux $\Phi$.  
Using standard flux freezing arguments (zero resistance, 
infinite conductivity, the magnetic flux $\Phi = B R^2$ = 
{\sl constant}), we obtain 
\be 
B_{IGM} = B_{\rm gal} \Bigl( {\rho_{\rm gal} \over \rho_{IG} } 
\Bigr)^{-2/3} (1 + z)^{-2} \sim 10^{-9} \, {\rm G} \, , 
\label{eq:fluxfreeze} 
\ee 
where $\rho_{\rm gal}$ is the typical density of the galaxy 
and $\rho_{IG}$ is the density of the IGM. The factor $(1+z)^2$ 
$\approx$ 16 takes into account the expansion of the universe 
since galaxies formed. 

[2] Far from the galaxy, the leading order term in the multipole 
expansion for the magnetic field strength is the dipole term, which 
decreases with radius as $r^{-3}$. Between galaxies, the magnetic 
dipole term thus contributes a characteristic field strength 
\be
B_{IGM} = B_{\rm gal} \Bigl( {\rho_{IG} \over \rho_{\rm gal} } 
\Bigr) \sim 10^{-12} \, {\rm G} \, . 
\label{eq:dipole}
\ee 
If $B_{IGM}$ is much smaller than this fiducial value, the galactic 
field geometry must have a rather special form so that the dipole 
term vanishes (or is highly suppressed). 

We note that other effects can increase the magnetic field 
strengths between galaxies beyond the simple estimates given here. 
For example, galactic scale winds can drag magnetic field lines 
into the intergalactic medium and thereby enhance the field 
strength of the IGM (Kronberg \& Lesch 1997). Many other 
magnetohydrodynamical effects can also take place, including 
field diffusion, reconnection, and dynamo activity 
(see, e.g., Shu 1992). 

Observations of magnetic fields are roughly consistent with the
estimates given above.  Although the magnetic field in our galaxy is
relatively well observed (e.g., Heiles 1976), magnetic fields are just
now being studied in external galaxies and the IGM (see, e.g., the
reviews of Kronberg 1994 and Biermann 1997).  Measurements of magnetic
fields in other galaxies suggest that our galaxy is fairly typical and
that field strengths of $B \approx 3 - 10$ $\mu$G are the norm.
Within galaxy clusters, magnetic fields can now be measured (e.g., Kim
et al. 1990) and the typical field strength in the central regions is
$B \sim 2$ $\mu$G.  Furthermore, the characteristic length for field
reversal is 13--40 kpc or perhaps even smaller (see Feretti et al. 1995). 
If we scale these values to larger size scales using a flux freezing
argument, the estimated IGM field strength is comparable to that of
equation [\ref{eq:fluxfreeze}].  Unfortunately, however, very little
data exist to constrain magnetic fields in the less dense regions of
the universe, i.e., outside the core regions of clusters (see Ensslin
et al. 1997 and Biermann 1997).  In one case, outside the Coma
cluster, the magnetic field strength has been estimated to be as large
as $10^{-7}$ G (Kim et al. 1989).  In addition, across cosmological
distances, there is an upper limit on the magnetic field strength,
$B_{IGM} < 10^{-9}$ G, which is estimated from rotation measure data
along the line of sight to quasars and modeling (Kronberg 1994; see
also Vall{\'e}e 1990).

Nearly independent of their origin, cosmic magnetic fields will be
pulled along by baryonic matter and will thus be tied to galaxies. 
Using this idea, models of the intergalactic magnetic fields are 
now being proposed (see, e.g., Biermann, Kang, \& Ryu 1996; 
Kronberg 1996; Kronberg \& Lesch 1997). This work shows that magnetic 
fields lines can in principle connect galaxies to each other and 
that the magnetic field strength can lie in the range $10^{-7}$ 
$-$ $10^{-10}$ G, consistent with the simple estimates given above. 
Notice also that the field lines between galaxies will tend to 
straighten out due to both magnetic tension and the expansion 
of the universe. 

To summarize, we take the magnetic field strength in the IGM 
to lie in the range $10^{-12}$ G  $< B_{IGM} < 10^{-7}$ G. 
The lower limit arises because of the dipole contribution from 
galaxies. The upper limit is indicated both by observations and 
by elementary physical considerations. 

We now estimate mean free paths for cosmic rays in the IGM. For most
of this paper, we use the limiting case in which the mean free path
$\ell$ is the typical distance between galaxies, i.e., $\ell \approx$
1 Mpc.  This choice provides an optimistic but plausible case for the
transport of low energy cosmic rays.  This limit is realized, for
example, if the magnetic field lines are straight between galaxies and
no other effects impede the propagation of particles.  Then the
intergalactic field lines are tied to the galactic field lines, so the
coherence length of the field can be as large as the mean separation
between galaxies.  However, galaxies are randomly oriented and it is
thus highly unlikely that the magnetic field reversal length scale is
much larger than 1 Mpc.  Notice that the exact value of the magnetic
field strength is not important in this case.

We also consider the opposite limit in which the magnetic field is 
extremely tangled and the cosmic rays are well coupled to the field.  
In this case, the effective mean free path $\ell$ is given by the 
magnetic gyro radius $r_B$, i.e., 
\be 
\ell = \lambda r_B = \lambda {E \over q B} \, = 1 \, {\rm pc} \,
\Bigl( {E \over 1 {\rm GeV} } \Bigr) 
\Bigl( {B \over 10^{-12} {\rm G} } \Bigr)^{-1} \, \lambda \, , 
\label{eq:gyro} 
\ee 
where $E$ is the energy of the particle, $q$ is the charge, and $B$ 
is the magnetic field strength. The dimensionless enhancement factor 
$\lambda$ takes into account the possibility that the mean free path 
can be somewhat longer than the magnetic gyro radius. 

For typical cosmic ray energies, $E$ $\sim 1$ GeV/n, where $n$ is 
the number of nucleons, the magnetic gyro radius is a factor of  
$\sim 10^6$ smaller than the most probable distance between galaxies.
Thus, to estimate the maximum possible influence of extragalactic
cosmic rays, we use the distance between galaxies $\ell \sim 1$ Mpc as
the optimal mean free path.  However, cosmic rays with energies larger
than about $E \sim 10^6$ $(B/10^{-12} {\rm G})$ GeV have magnetic gyro
radii larger than 1 Mpc; these high energy cosmic rays experience a
completely tangled magnetic field and thus have a mean free path given
by equation [\ref{eq:gyro}].

Given the mean free path for both high and low energy cosmic rays,
we can now estimate how far these particles travel in a given time period.
For diffusion in a uniform medium (e.g., the IGM), the distance traveled
in a time interval $t$ is roughly given by the diffusion length $R_0$,
which we write in the form
\be
R_0 = \bigl[ D \, t \bigr]^{1/2} =
\bigl[ c \, t \, \ell / 3 \bigr]^{1/2} \, = 32 \, {\rm Mpc}
\, (\ell/1{\rm Mpc})^{1/2} \, (t/10 {\rm Gyr})^{1/2} \, ,
\label{eq:dlength}
\ee
where $c$ is the particle speed (the speed of light), and $\ell$ is the
mean free path.  Thus, in the age of the universe, these particles have 
traveled a distance much smaller than the horizon size, 
$\sim 3000 h^{-1}$ Mpc. 

\section{A Model for Cosmic Ray Diffusion}	\label{sec:model} 

In order for cosmic rays from large distances to enter our Galaxy,  
they must overcome two hurdles: [1] The particles must first diffuse 
through the IGM to reach the general vicinity of our Galaxy. 
[2] The particles must enter the Galaxy itself by overcoming the
barriers provided by magnetic fields, galactic winds, and other
effects. In this section, we develop a model for cosmic ray diffusion
and include the effects of the fractional accessibility to the galaxy.

This model is based on the simplified picture of intergalactic
magnetic fields as discussed in \S 2.  It is important to keep in 
mind that cosmic ray propagation depends on the particle energy. 
In this picture, low energy cosmic rays follow field lines from galaxy
to galaxy with a mean free path of $\sim 1$ Mpc; 
most of this section deals with this low energy case. 
On the other hand, high energy cosmic rays perform a random walk that
is almost independent of the galaxies; we consider the high energy 
limit in \S 3.6.  We also note that other magnetic field configurations 
are possible.  However, this present formulation is rather robust in 
that it can be applied to many different specific models for magnetic
field configurations and cosmic ray propagation, provided only that
the evolution is diffusive.

\subsection{A Self-Similar Diffusion Model} 

The cosmic ray flux from any given galaxy has a limited sphere of 
influence because of the relative difficulty for particles to diffuse 
through the IGM. The number density and the flux of cosmic rays can 
be described by a simple similarity solution. We begin by writing 
the diffusion equation in the form 
\be 
{\partial n \over \partial t} = D \nabla^2 n  - 
\Lambda n \, , 
\label{eq:diffuse}
\ee 
where $D = c \ell/3$ is the diffusion constant and the parameter 
$\Lambda$ accounts for the destruction of cosmic rays.  The solution 
for the number density of cosmic rays can be written in the form  
\be
n(r,t) = t^{-1/2} \, {\rm e}^{-\Lambda t} \, f(\xi) \, , 
\label{eq:form}
\ee 
where the power-law index $1/2$ is chosen so that the galaxy has a 
constant luminosity $L_{CR}$ of cosmic rays. The similarity variable 
$\xi$ is defined by $\xi \equiv r/R_0$, where $R_0$ is a characteristic 
length scale $R_0$ $\equiv$ $[c t \ell / 3 ]^{1/2}$ $\approx 32$ Mpc 
$\, (\ell/1{\rm Mpc})^{1/2} \, (t/10 {\rm Gyr})^{1/2} \, $ 
(see equation [\ref{eq:dlength}]). 

Notice that we neglect the expansion of the universe in this simple
treatment of the problem. This treatment is justified because the
effective diffusion length (see below) is much smaller than the
horizon size.  In addition, the inclusion of the expansion of the
universe introduces additional assumptions (e.g., an open versus
closed universe). 

With the above definitions, the resulting ordinary differential 
equation for the reduced diffusion field $f(\xi)$ takes 
the simple form 
\be 
f_{\xi \xi} + {2 \over \xi} f_\xi + {\xi \over 2} f_\xi + 
{1 \over 2} f = 0 \, , 
\label{eq:diffeq} 
\ee 
where subscripts denote differentiation. The relevant boundary 
conditions for this problem are [1] the cosmic ray luminosity 
approaches a constant at the origin ($f \to 1/\xi$ as $\xi \to 0$), 
and [2] the cosmic ray flux outward through any given spherical 
shell vanishes at spatial infinity ($\xi^2 f \to 0$ as $\xi \to \infty$). 
With these boundary conditions, equation [\ref{eq:diffeq}] 
has the solution 
\be 
f(\xi) = f_0 { 1 - {\rm erf} (\xi/2) \over \xi } \, , 
\label{eq:fsolution}
\ee 
where the constant $f_0$ is determined by the cosmic ray luminosity 
of the galaxy and erf$(z)$ is the error function. The flux $\cal F$ 
of cosmic rays at a distance $r$ from the galaxy is given by 
${\cal F}$ = $- D {\partial n / \partial r}$. The cosmic ray 
luminosity $L_{CR}$ of the galaxy is 
\be
L_{CR} = \lim_{\xi \to 0} 4 \pi r^2 {\cal F} = 
\lim_{\xi \to 0} - 4 \pi D r^2 {\partial n \over \partial r} 
= 4 \pi f_0 (\ell c / 3)^{3/2} \, . 
\ee 
It is useful to combine these results to write the cosmic ray 
flux in the form 
\be
{\cal F} = {L_{CR} \over 4 \pi r^2} \, g(\xi) 
\, {\rm e}^{-\Lambda t} \qquad {\rm where} \qquad 
g(\xi) \equiv \Bigl\{ 1 - {\rm erf}(\xi/2) + 
\xi \pi^{-1/2} \exp[-\xi^2/4] \Bigr\} \, . 
\ee
The function $g(\xi)$ thus encapsulates the departure of the 
cosmic ray flux from the naive result ${\cal F} = L_{CR}/4 \pi r^2$ 
which applies in the limit of no diffusion ($\ell \to \infty$). 
Note that this diffusive argument does not include energy loss 
of the particles from adiabatic cooling. 

\subsection{Fractional Accessibility Argument}

We must take into account the possible destruction of cosmic rays as
they travel through the universe.  We have proposed a model in which
cosmic rays propagate from galaxy to galaxy and finally reach our
galaxy.  Each galaxy along the way has some accessibility fraction
$x$, i.e., the fraction of incident cosmic rays that actually enter
the galaxy.  Once inside a galaxy, cosmic rays bounce around for a
time $\tau_{esc}$, the escape time. The cosmic rays have some chance 
of interacting and being destroyed with a characteristic time scale 
$\tau_{int}$.  As a result, the fraction $f_1$ of cosmic rays that
remain after each galactic visit (each step of the random walk) is
given by 
\be 
f_1 = (1 - x) + {x \over 1 + \tau_{esc}/\tau_{int}}
\equiv 1 - \alpha x \qquad {\rm where} \qquad \alpha 
\equiv {\tau_{esc} \over \tau_{int} + \tau_{esc} } \, .  
\ee

We note that the destruction parameter $\alpha$ is energy dependent. 
For relatively low energies ($E$ $< 10$ GeV/n), the interaction time 
and the escape time are comparable, 
$\tau_{int} \sim \tau_{esc} \sim 10^7$ yr (Ormes \& Freier 1978), 
and the parameter $\alpha \approx 1/2$.  These results are based 
on solid experimental measurements. The escape time $\tau_{esc}$ 
is based on the abundance of the radioactive species $^{10}$B 
and the interaction time $\tau_{int}$ is based on the well 
measured secondary to primary ratio (B/C). At very high energies, 
$\tau_{esc}$ and $\tau_{int}$ are not well determined experimentally. 
For energies $E >$ 10 TeV/n, element separation becomes difficult 
and the secondary to primary ratio (B/C) becomes uncertain. 
However, one would expect that the escape time becomes comparable 
to the light crossing time of the galaxy, 
$\tau_{esc} \sim 3 \times 10^4$ yr, and 
$\alpha$ approaches a smaller value, at least $1/300$. 

As an aside, we note that an escape time $\tau_{esc} \sim 10^7$ yr,
combined with the coherence length of the galactic magnetic field
$\ell \sim 300$ pc, implies an effective diffusion length of
$L_D = [c \tau_{esc} \ell ]^{1/2} \approx 30$ kpc, a length
scale comparable to the size of the galaxy.  It is thus plausible
that a diffusion model applies to the escape of cosmic rays from
the galaxy.  However, since we know the escape time $\tau_{esc}$
from experimental considerations, we need not use such a model
for this paper.

On average, each cosmic ray will interact with 
$N \approx c \tbar /\ell$ galaxies, where the average time 
$\tbar$ since the cosmic ray was emitted is half the age of 
the galaxy and hence $N \approx 1500$ $(\ell/ 1 {\rm Mpc})^{-1}$. 
After $N$ interactions, the total remaining fraction $f_N$ of cosmic 
rays is $f_N$ = $f_1^N$ = $(1 - \alpha x)^N$. In the large $N$ limit, 
this expression approaches the form $f_N \approx \exp[-\alpha N x]$ 
and we make the identification $\Lambda t = \alpha x N$ to specify the
destruction rate $\Lambda$ in terms of the other parameters of the
problem, i.e., $\Lambda = \alpha x c / 2 \ell$.

This interaction loss from the cosmic ray flux sets up an
accessibility problem. In order for cosmic rays to survive the
diffusion process, the fractional accessibility $x$ must be very
small. If the fraction $x$ is small, however, the particles have
little chance of entering our own galaxy.  These two competing 
effects imply that the flux of cosmic rays into our Galaxy is 
proportional to the ``destruction function'' $F_D$ defined by 
\be
F_D (x) = x (1 - \alpha x)^N \, . 
\ee
Here, the factor $x$ is the probability that a cosmic ray will enter 
our Galaxy and the factor $(1 - \alpha x)^N$ is the probability of 
surviving other galaxies en route.  Hence, the function $F_D (x)$ 
represents the fraction of the total possible cosmic ray flux that 
can enter our Galaxy.  The destruction function has a maximum value 
$F_{\rm max}$ =
$\alpha^{-1} (N+1)^{-1} [ N/N+1 ]^N$ $\approx$
$({\rm e} \alpha N)^{-1}$ at the critical value
$x$ = $[\alpha (1+N)]^{-1}$.
For typical values ($\alpha = 1/2$; $\ell$ = 1 Mpc), we find
$F_{\rm max} \approx 5 \times 10^{-4}$ at $x \sim 10^{-3}$.
We obtain essentially the same result in the continuum limit 
using the form
\be
F_D (x) \equiv x {\rm e}^{-\Lambda t} = x {\rm e}^{-\alpha x N} 
\ee
for the destruction function. Notice that the mean number $N$ of
galaxies in the path of a cosmic ray depends on the mean free path,
i.e., $N \approx 1500 (\ell/1 {\rm Mpc})^{-1}$.  For most of
parameter space, the function $F_D$ is quite small and is sharply
peaked about its maximum value. In particular, $F_D \to 0$
for both limiting cases $x \to 0$ and $x \to 1$. 
 
The fractional accessibility $x$ can in principle be calculated.  In
order to enter a galaxy, cosmic rays must propagate through galactic
winds and any other inhibiting factors.  One such calculation (Ahlen
et al. 1982) uses a planar diffusion model to represent the disk of
the galaxy; these authors find that the accessibility parameter $x$ 
is close to 1/10 at GeV energies and approaches unity at high energies. 
Subsequent work using a spherical diffusion model found comparable
results for $x$.  The galaxy is expected to behave in a manner
intermediate between the planar and the spherical models.  In the
spherical diffusion case, however, some adiabatic losses occur; these
losses are analogous to the case of the solar wind. For the case of
spherical diffusion, adiabatic losses shift the cosmic ray spectrum in
energy by perhaps several GeV. Because of this effect, the cosmic rays
one observes at a given energy represent the IGM spectrum at slightly
higher energies. Since the flux of cosmic rays decreases rapidly with
energy, the net effect of these losses is to reduce the total flux at
a given energy.  This effect is of course most important at low
energies comparable to the energy shift (perhaps several GeV).  
In this paper, however, we are interested in finding the largest
possible cosmic ray flux from other galaxies, so we do not include
these losses.

The energy shift of several GeV can be roughly approximated. The energy
scale at which modulation becomes important can be estimated by the
condition $V R/D = 1$, where $V$ $\sim$ 10 km/s is the speed of the
galactic wind, $R$ $\sim$ 10 kpc is the size of the galaxy, and 
$D = c \ell/3$ is the diffusion coefficient (see Ahlen et al. 1984 for
further discussion of all of these points).  Using this relation, 
and scaling to the case of the solar wind, we find an energy shift 
of about 10 GeV.  

\subsection{Extragalactic Cosmic Rays} 

We now estimate the total flux of cosmic rays that are emitted by 
external galaxies and absorbed by our galaxy.  The total flux 
${\cal F}_T$ of extragalactic cosmic rays that impinge upon our 
galaxy is given by the integral 
\be
{\cal F}_T = L_{CR} \, n_{\rm gal} \, R_0 \, {\rm e}^{-\Lambda t} 
\, \int_0^\infty g(\xi) d\xi = 4 \pi^{-1/2} \, 
L_{CR} \, n_{\rm gal} \, R_0 \, {\rm e}^{-\Lambda t} \, , 
\ee
where $n_{\rm gal}$ is the number density of galaxies. Since this 
integral only has support in the local portion of the universe, we 
need not consider the curvature of the universe in this evaluation.  
We can now estimate the fraction of cosmic rays within our galaxy 
that have an extragalactic origin.  The rate of absorption of 
extragalactic cosmic rays by our galaxy is given by 
\be
L_X = 2 \pi R_D^2 x {\cal F}_T \, , 
\ee
where $x$ is the fraction of the incident cosmic rays that enter 
the galaxy (we assume that the galaxy can be modeled as a disk 
with radius $R_D \sim 15$ kpc). Within our Galaxy, the fraction 
of cosmic rays $\chi$ that have an extragalactic origin becomes 
\be
\chi \equiv {L_X \over L_{CR} + L_X } \approx {L_X \over L_{CR} } = 
8 \pi^{1/2} [R_D^2 R_0 n_{\rm gal} ] \, x {\rm e}^{-\Lambda t} 
\approx 0.1 (\ell/1 {\rm Mpc})^{1/2} \, x {\rm e}^{-\Lambda t} \, . 
\label{eq:xfraction} 
\ee 
The result thus depends on the ``destruction function'' $F_D$ defined
above.  Since the destruction function has a maximum value of 
$F_D = ({\rm e} \alpha N)^{-1} \sim 5 \times 10^{-4}$, the maximum 
expected fractional abundance of extragalactic cosmic rays is
similarly small, i.e., $\chi = L_X/L_{CR} < 5 \times 10^{-5}$ 
(for a mean free path $\ell$ = 1 Mpc).  For most of parameter space, 
the fraction of extragalactic cosmic rays is extremely small. 

\subsection{Detecting A Possible Anti-Matter Signal}  

Ever since antiprotons were discovered (Chamberlain et al. 1955), 
searches for extragalactic anti-matter have steadily improved.
Antiprotons of secondary origin have recently been found (e.g.,
Yoshimura et al. 1995; Mitchell et al. 1996), but heavier anti-matter
which could originate in a cosmic anti-matter domain (primary
anti-matter) has not been detected. The experimental situation prior
to 1980 is summarized in Ahlen et al. (1982).  Subsequent limits on
${\overline{\rm He}}$/He using balloon experiments have reached the 
$8 \times 10^{-6}$ level (Ormes et al. 1997) and may ultimately reach
$10^{-7}$ with the development of long duration ballooning. 
A planned space experiment, the Alpha Matter Spectrometer (AMS), 
may conceivably reach the $10^{-8}$ level (see, e.g., Ahlen et
al. 1994).  It is thus worthwhile to consider the possible importance
of obtaining even tighter bounds on extragalactic anti-matter.  This
discussion assumes low energy cosmic rays, the regime accessible by
future experiments.  Furthermore, this treatment attempts to be
optimistic in the sense that we try to find the largest possible
anti-matter signal within a plausible class of models.

To consider cosmic rays emitted by distant (hypothetical) 
anti-galaxies, we find the fraction ${\cal R}(a)$ of cosmic 
rays originating at distances greater than some scale $a$, 
\be
{\cal R} (a) \equiv {\int_{\xi_a}^\infty g(\xi) d\xi 
\over \int_0^\infty g(\xi) d\xi } = \exp[-\xi_a^2/4] - 
{ \pi^{1/2} \over 4 } \, \xi_a \, [ 1 - {\rm erf} (\xi_a/2) ] 
\, \approx {1 \over 2} \exp[-\xi_a^2/4] \, , 
\ee 
where $\xi_a = \xi(a) = a/R_0$. The final equality evaluates this 
ratio in the asymptotic limit ($\xi_a \to \infty$).  For large 
distances $a \gg R_0$, the ratio $\cal R$ decays like a gaussian 
and hence the volume of the universe that produces cosmic rays 
accessible to our galaxy has a radius of $\sim 2 R_0 \approx 64$ Mpc. 
The fractional abundance $\cal A$ of extragalactic anti-matter 
is given by the product ${\cal A} = \chi {\cal R} f_A$, where $f_A$ 
is the fraction of anti-galaxies (for a baryon symmetric universe, 
$f_A$ = 1/2). The expected fraction $\cal A$ of anti-matter 
in the cosmic ray flux can be written 
\be
{\cal A} = 0.025 \, {\tilde \ell}^{1/2} x 
\exp[ - (750 x + 244 {\tilde a}^2)/ {\tilde \ell} ] \, , 
\label{eq:antifraction} 
\ee
where the scaled variable ${\tilde \ell}$ is the mean free path 
in units of Mpc and ${\tilde a}$ is the distance to the nearest 
anti-galaxy in units of 1000 Mpc.  In Figure 1, we have plotted the
expected fractional abundance $\cal A$ of extragalactic anti-matter 
in the cosmic ray flux as a function of distance $a$ to the nearest 
anti-matter domain. 

Current experiments place rather tight constraints on the distance
scale $a$ to the nearest anti-matter domain (from Steigman 1976 to
Cohen 1996).  The strongest limits arise from considering the expected
gamma ray flux from matter/anti-matter annihilations.  The absence of
copious gamma rays from galaxy clusters implies that $a >$ 40 Mpc
(e.g., Peebles 1993).  Recent calculations (Cohen 1996) indicate that
hypothetical anti-galaxies must be no closer than $a \approx 1500$ Mpc
(see also Dudarewicz \& Wolfendale 1994); 
this lower limit implies that the ratio ${\cal R} \sim 10^{-239}$ 
for mean free path ${\tilde \ell}$ = 1 and for the optimal value of $x$. 
An independent calculation using the Sunyaev-Zel'dovich effect rules
out anti-matter domains to a distance scale of $a \sim 200$ Mpc
(Cohen, private communication).  Finally, additional constraints 
can arise from annihilation signatures in the cosmic background 
radiation (see Kinney, Kolb, \& Turner 1997). 

To illustrate the difficulty associated with anti-matter domains, we
present the following order of magnitude argument.  We consider domain
regions with size scale $\lambda \approx a$ and luminosity $L_D$ in
gamma rays from matter/anti-matter annihilation in the overlap
regions.  The total observed gamma ray flux is 
$F_\gamma \approx L_D (c t_0)^4 \, \lambda^{-6}$, 
integrated from the nearest domain boundary (at distance $\lambda$)
out to the redshift at which the domains enter the horizon.  The
luminosity $L_D = g {\dot N_N}$, where $g \sim 4$ is the number of
photons per annihilation and the annihilation rate per domain 
$\dot N_N$ = $\epsilon \lambda_{100}^3 \, 5 \times 10^{55}$ 
${\rm s}^{-1}$. The efficiency $\epsilon$ is the fraction of the 
domain that annihilates during the age of the universe and
$\lambda_{100}$ is the domain scale in units of 100 Mpc.  By comparing
this expected gamma ray flux to the observed background flux, we
obtain the constraint $\lambda_{100}^3 \ge 10^{14} \epsilon$.  To
estimate the efficiency $\epsilon$, we assume spherical domains for
which only the outer layer (of thickness $\ell_T$) is available for
annihilation: $\epsilon = 3 \ell_T / \lambda$.  For cosmological
structures, the thickness $\ell_T$ should be of order the scale of
galaxy formation, $\ell_T \sim 1$ Mpc.  For this case, we obtain
$\epsilon \sim 0.03 \lambda_{100}^{-1}$ and hence a lower bound
$\lambda_{100} > 1320$ (larger than the horizon size and hence
unphysical).  To allow nearby anti-matter domains ($\lambda_{100} \sim
1$), the efficiency must be very small $\epsilon \sim 10^{-14}$, which
implies an unreasonably small domain thickness $\ell_T \sim 10^{12}$ cm
(only $\sim10$ stellar radii).

To evaluate the anti-matter fraction $\cal A$ for a given distance 
scale $a$, we first consider the maximal case by optimizing the 
function $\cal A$ with respect to the fractional accessibility $x$, 
i.e., ${\cal A}$ = $10^{-5} {\tilde \ell}^{3/2}$  
$\exp[ - 244 {\tilde a}^2/ {\tilde \ell} ]$.
Thus, for the representative values $\tilde \ell$ = 1 = $\tilde a$,
the maximum allowed anti-matter fraction is only ${\cal A} \sim
10^{-111}$. If we use a more typical value of the fractional
accessibility $x = 0.1$, the anti-matter abundance $\cal A$ will be
much smaller. Notice that the anti-matter fraction $\cal A$ has
exponential sensitivity to the fractional accessibility $x$ and has
gaussian sensitivity to the distance scale $a$ of anti-matter domains;
this extreme sensitivity to the input parameters cannot be
overemphasized (see Fig. 1).  Indeed, the situation is such that no
realistically conceivable improvement in experimental sensitivity 
to anti-matter could significantly increase the distance to which
putative anti-matter domains can be detected.

\subsection{The Large $\ell$ Limit}

For completeness and comparison, we consider the limit in which the
mean free path becomes very large, i.e., the limit of no diffusion.
This case is not expected to be realized in practice because any
magnetic fields in the universe cause particle propagation to deviate
from straight paths.  Nonetheless, it is instructive to consider this
limiting case. In this case, cosmic rays can travel straight to our
Galaxy and we must consider the curvature of the universe.  In this 
limit, the total extragalactic cosmic ray flux becomes 
\be
{\cal F}_T = L_{CR} \int n_{\rm gal} dr = 
L_{CR} c t_0 n_0 \bigl[ (1 + z)^{3/2} - 1 \bigr] \, , 
\ee
where we assume a spatially flat universe and $z \approx 3$ is 
the redshift at which galaxies begin to emit cosmic rays.  The 
ratio $\chi$ of extragalactic to galactic cosmic rays becomes 
\be
\chi = {L_X \over L_{CR} } = 2 \pi x R_D^2 c t_0 n_0 
\bigl[ (1 + z)^{3/2} - 1 \bigr] \, \approx 30 x . 
\ee 
Thus, unless the accessibility fraction $x$ is very small,
extragalactic cosmic rays must provide a significant fraction of 
the total flux in the no diffusion limit.  Notice that for cosmic 
rays with energies less than a few GeV, adiabatic losses will be 
significant and hence the flux of extragalactic cosmic rays will be
highly suppressed.  For cosmic rays with energies much higher than a
few GeV, adiabatic losses are small and the above estimate is valid. 

\subsection{The High Energy Limit} 

We also consider the limit of high energy cosmic rays with 
$E \gg 1$ GeV.  In this limit, the destruction parameter $\alpha$, 
the fractional accessibility $x$, and the mean free path $\ell$ 
obtain different values than in the opposite (low energy) limit. 
These high energy cosmic rays are not tightly bound to the galaxy 
and can escape on the light crossing time scale. At high energies, 
the destruction parameter becomes small, $\alpha \ge 1/300$, and the 
fractional accessibility $x$ approaches unity. In this limit, the 
fraction $\chi$ of extragalactic cosmic rays obtains the simple form 
\be
\chi = {L_X / L_{CR} } = 0.1 \, {\tilde \ell}^{1/2} 
{\rm e}^{-5/{\tilde \ell}} \, . 
\ee 
In the intermediate high energy regime, 1 GeV $\ll E \ll 10^6$ GeV, 
the mean free path $\tilde \ell \approx 1$, and the fraction of 
extragalactic particles approaches the constant value 
$\chi = L_X/L_{CR} \approx 7 \times 10^{-4}$. 

At larger energies, $E \gg 10^6$ GeV ($B/10^{-12}$ G), the mean free
path increases (see equation [\ref{eq:gyro}]) and the fraction of
extragalactic cosmic rays increases accordingly.  However, for
energies less than $\sim 10^{18}$ eV, the galactic magnetic field
scrambles the directions of extragalactic cosmic rays.  Thus, a window
exists for observing anisotropy in the extragalactic cosmic ray
signal, $10^{18} {\rm eV} < E < 3 \times 10^{19} {\rm eV}$, where the
upper limit (the GZK limit) arises from the interaction of cosmic rays 
with the cosmic microwave background (Greisen 1966; Zatsepin \& Kuzmin 
1966).  In this window, for $B_{IGM} = 10^{-9}$ G, the fraction $\chi$ 
lies in the range $7 \times 10^{-4} < \chi < 0.46$ according to this
model; these $\chi$ values provide upper limits on the anisotropy. 
Future experimental searches for extragalactic cosmic rays should 
thus concentrate on the high energy regime (for the current 
experimental situation regarding arrival directions of high energy 
cosmic rays, see, e.g., Stanev et al. 1995; Watson 1996). 

The validity of a diffusion model requires that the cosmic rays
experience a large number of scattering events on their way to our
galaxy. In the high energy limit considered here, the number $N$ 
of scatterings depends on the magnetic field strength, where 
$N = c t / 2 \ell$ and $\ell$ is now given by the magnetic gyro
radius.  For IGM fields with large but representative field strengths, 
$B_{IGM} = 10^{-9}$ G, the number $N \approx 50$ at the highest
energies, the GZK limit. The diffusion approximation thus remains
marginally satisfied.  However, for a weaker field, $B_{IGM} =
10^{-12}$ G, the number $N$ of scatterings becomes of order unity for
particle energies greater than $E \sim 10^{18}$ eV.  In this case, 
for relatively weak intergalactic magnetic fields, the considerations 
of the previous subsection apply. 

\subsection{Long Term Evolution} 

We have shown that any galaxy will have a rather small sphere of
influence from its cosmic ray output.  In particular, this sphere of
influence (given by $R_0 \sim 32$ Mpc ($\ell$/1Mpc)$^{1/2}$
($t$/10Gyr)$^{1/2}$) is much smaller than the horizon size scale
($\sim 3000 {\rm h}^{-1}$ Mpc), but much larger than the
characteristic distance between galaxies ($\sim 1$ Mpc). In order 
to gain further understanding of this problem, we consider the future
evolution of cosmic ray diffusion in the universe for time scales that
exceed the current Hubble time (see, e.g., Adams \& Laughlin 1997 for
a recent review of long term effects in the universe).  The co-moving
diffusion length can be written in the form 
\be
{\widetilde R}_0 = {R_0 \over R(t) } =
\Bigl[ {c t \ell_0 \over 3 R(t) } \Bigr]^{1/2} \, ,
\ee
where $R(t)$ is the scale factor of the universe.

For a spatially flat universe, $R(t) \sim t^{2/3}$, and hence the
co-moving diffusion length is a slowly growing function of time.
Thus, the cosmic ray flux from a given galaxy will gradually influence
galaxies of increasing distances. However, the co-moving diffusion
length grows much slower than the horizon.  As a result, the sphere of
influence of any given galaxy will correspond to a decreasing fraction
of the total volume of the observable universe as time proceeds.  If
the universe is open, then the scale factor approaches the form $R(t)
\sim t$ in the relatively ``near'' future. In this case, the co-moving
diffusion length approaches a constant value asymptotically in time.
For completeness, we note that additional dynamical evolution of the
magnetic fields will also take place.

\section{Summary} \label{sec:sum} 

We have presented a simple analytic model for the diffusion of cosmic
rays through intergalactic space.  This model clearly elucidates the
difficulty faced by particles propagating large distances through the
IGM. The results of this model are summarized below:

[1] The magnetic field strength estimated for the IGM lies in the
range $B_{IGM} = 10^{-12} - 10^{-7}$ G.  Low energy cosmic rays
($E < 10^6$ GeV) tend to follow the magnetic field lines. 
As a reasonable model for long distance particle transport, 
we take the cosmic ray mean free path to be comparable to the 
mean separation of galaxies, $\ell \sim 1$ Mpc. We take high 
energy cosmic rays to have a mean free path given by the 
magnetic gyro radius. 

[2] The cosmic ray output of a galaxy has a sphere of influence
with radius $R_0 = [c t \ell/3]^{1/2} \approx 32$ Mpc
($\ell$/1Mpc)$^{1/2}$ ($t$/10 Gyr)$^{1/2}$.
The time scale for particles to travel large distances
($r \gg R_0 \sim 32$ Mpc) through the IGM is thus much
longer than the current age of the universe.

[3] We have demonstrated an accessibility problem for low energy
extragalactic cosmic rays. In this model, cosmic rays diffuse through
many different galaxies on the way to our Galaxy. In each galaxy,
cosmic rays have some chance of being destroyed. In order for cosmic
rays to survive the diffusion process, the fractional accessibility
$x$ must be small; if the accessibility $x$ is small, however, cosmic
rays have little chance of entering our Galaxy. This compromise sets
up a maximum survival fraction of $F_{\rm max} \sim 5 \times 10^{-4}$.

[4] The fractional abundance of low energy extragalactic cosmic rays 
is extremely small for this model (eq. [\ref{eq:xfraction}]). The 
abundance of extragalactic cosmic rays is exponentially suppressed 
by the fractional accessibility effect described in item [3].

[5] Since hypothetical galaxies made of anti-matter must be fairly
distant ($a > 1000$ Mpc $\gg R_0$), the abundance of anti-matter in 
the cosmic ray flux corresponds to the gaussian tail of the distribution.  
As a result, the fractional abundance of anti-matter is expected to 
be small (eq. [\ref{eq:antifraction}] and Fig. 1) even for the extreme 
case of a baryon symmetric universe. Although anti-matter domains in 
the universe remain an interesting possibility, it must be realized 
that cosmic rays do not provide an effective search method. 

[6] If cosmic rays propagate freely rather than diffuse, the
fractional abundance of extragalactic cosmic rays would be much
higher, as large as $30x$, where $x$ is the fractional accessibility.
This case is essentially ruled out by existing experimental
constraints, but more definitive data will be forthcoming.

[7] We have shown that a window exists for observing cosmic 
ray anisotropy at high energies in the range $10^{18} {\rm eV}$ 
$< E <$ $3 \times 10^{19}$ eV (see section 3.6). 

[8] This formulation is in some sense more robust than its 
derivation because it has been posed parametrically. 
In particular, it can be applied to many different 
specific models for cosmic ray propagation and magnetic field 
configurations, provided that the evolution is diffusive.

\acknowledgements

We would like to thank A. Cohen, J. Beatty, K. Green, P. Kronberg,
M. Lemoine, J. Matthews, E. Parker, R. Rosner, G. Sigl, S. Swordy and
J. Truran for useful discussions.  We also thank an anonymous referee
for many useful comments that improved the paper. This work was
supported by a NSF Young Investigator Award, NSF PHY 9406745, NASA
Grant No.~NAG~5-2869, DOE Theory Grant, DOE Grant
No. DE-FG02-95ER40899, and by funds from the Physics 
Department at the University of Michigan.

\newpage 
\begin{center}
\large Figure Captions
\end{center}

\figcaption{
Expected fractional abundance $\cal A$ of extragalactic anti-matter 
in the cosmic ray flux as a function of distance $a$ to the nearest 
anti-matter domain.  The curves at the left show $\cal A$ for mean
free path $\ell$ = 1 Mpc and varying values of the fractional
accessibility $x$: top curve uses optimal value $x = 1.33 \times
10^{-3}$; middle curve uses $x = 10^{-4}$; bottom curve uses $x =
10^{-2}$.  The curve using the most likely value $x = 0.1$ has
abundance values $\cal A$ less than $10^{-20}$ over the entire
distance range shown.  The current best experimental limits are shown
as solid horizontal lines, whereas the sensitivity of proposed
measurements are indicated with horizontal dotted lines.  The 
regions of parameter space excluded by $\gamma$-ray flux 
considerations (Cohen 1996) and the horizon distance are 
also indicated.  
\label{figF}}

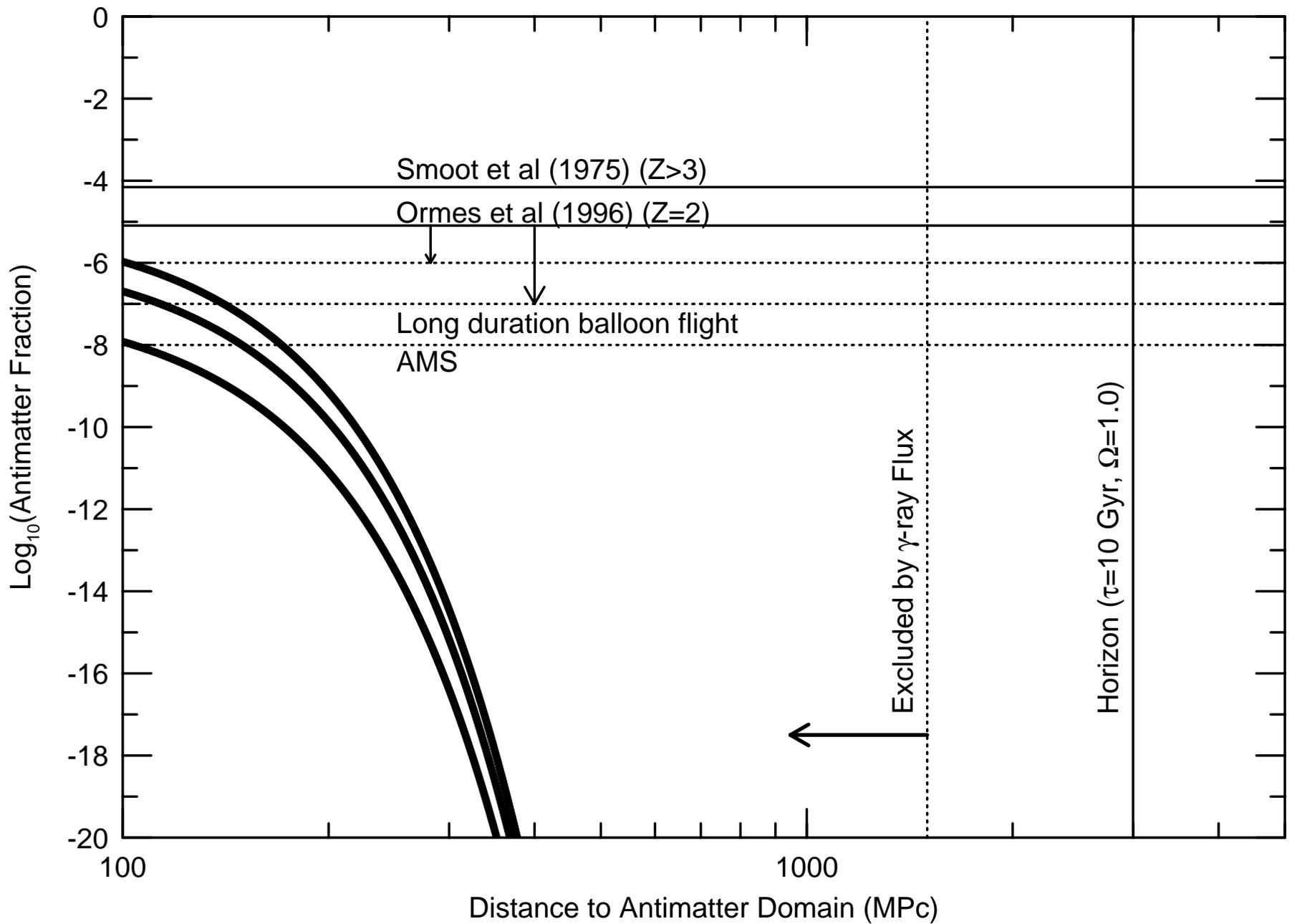
\end{document}